\documentclass[aps,prb,reprint,groupedaddress,amsmath,amssymb,twocolumn]{revtex4}

\usepackage{mathrsfs}
\usepackage{graphicx}
\usepackage{color}
\usepackage{cases}
\usepackage{array}
\usepackage{float}
\usepackage{hyperref}

\begin{document}

\title{Spatial inversion symmetry breaking of vortex current in biased-ladder superfluid}

\author{Weijie Huang$^{1}$ and Yao Yao$^{1,2}$\footnote{Electronic address:~\url{yaoyao2016@scut.edu.cn}}}

\address{$^1$ Department of Physics, South China University of Technology, Guangzhou 510640, China\\
$^2$ State Key Laboratory of Luminescent Materials and Devices, South China University of Technology, Guangzhou 510640, China}

\date{\today}

\begin{abstract}
We investigate the quench dynamics of interacting bosons on a two-leg ladder in presence of a uniform Abelian gauge field. The model hosts a variety of emergent quantum phases, and we focus on the superfluid biased-ladder phase breaking the $Z_{2}$ symmetry of two legs. We observe an asymmetric spreading of vortex current and particle density, i.e., the current behaves particle-like on the right and wave-like on the left, indicating spontaneous breaking of the spatial inversion symmetry. By decreasing the repulsion strength, it is found the particle-like current is more robust than the wave-like one. The evolution of entanglement entropy manifests logarithmic growth with time suggesting many-body localization matters.
\end{abstract}

\maketitle

Lattice gauge theory (LGT), an equivalent formalism of gauge field, has manifested advantages in comprehending many effects in condensed matter physics \cite{ichinose2014lattice}. Enormous progresses of quantum computations in the last decade have then stimulated intense research activities to explore the possibility of encompassing LGTs \cite{wiese2014quantum,zohar2016quantum,brunner2014bell, degen2017quantum, georgescu2014quantum,preskill2018quantum}. Currently, proposals have been made in a couple of platforms, including cold atoms in optical lattices, trapped ions, Rydberg atoms, and superconducting qubits \cite{banerjee2013Atomic, barrett2013Simulating, glaetzle2014Quantum, glaetzle2015Designing, hauke2013Quantum, homeier2023Realistic, mezzacapo2015NonAbelian, tagliacozzo2013Simulation,banerjee2012atomic, buchler2005atomic, byrnes2006simulating, tagliacozzo2013optical, zohar2011confinement,guan2020synthetic}, and proof-of-principle experimental realizations of lattice gauge fields have been conducted \cite{kokail2019selfverifying, martinez2016realtime, mil2020scalable, schweizer2019floquet, yang2020observation,surace2020lattice,tai2017microscopy}.

A specific type of gauge fields, known as background (or static) gauge field \cite{banuls2020simulating,dalmonte2016lattice, zohar2022quantum,ozawa2019topological}, has drawn significant attention \cite{aidelsburger2013realization, alba2011seeing, boada2010simulation, goldman2009ultracold, jaksch2003creation, jaksch2005cold, juzeliunas2008double, miyake2013realizing, mueller2004artificial, osterloh2005cold, paredes2008minimum, ruseckas2005nonabelian, sorensen2005fractional}. This field, together with proper interactions, generate an extensive collection of many-body phases with remarkable properties, such as superconductivity and Mott insulator. By synthesizing spatial dimensions with gauge, one-dimensional quantum simulators provide convenient realizations of higher-dimensional quantum models \cite{boada2012quantum,celi2014synthetic,ozawa2019topological}, such as the Harper-Hofstadter model \cite{celi2014synthetic,aidelsburger2013realization, goldman2009ultracold, jaksch2003creation, miyake2013realizing, osterloh2005cold}. These studies have also paved ways for investigating other scenarios including both bosonic and fermionic ladders under magnetic field, as the Hall effect has been calculated\cite{barbarino2015magnetic,buser2021probing,buser2022snapshotbased,calvanesestrinati2017laughlinlike,jaksch2005cold,petrescu2015chiral,taddia2017topological,sorensen2005fractional,greschner2019universal} and measured \cite{genkina2019imaging,mancini2015observation,zhou2022observation}. Research interests were further extended to the non-equilibrium dynamics by imposing a quantum quench \cite{banerjee2013Atomic,mancini2015observation,buser2020interacting,guan2020synthetic,mikhail2022resonant,tai2017microscopy}. The dynamics of a substantial microscopic particle was simulated, e.g., two repulsively interacting bosons on a real-space flux ladder \cite{tai2017microscopy}. The sensitivity of model parameters and initial states has been noticed in the short-time dynamics which is essential in experimental simulations \cite{mancini2015observation,buser2020interacting,tai2017microscopy,guan2020synthetic,mikhail2022resonant}.

As shown in a landmark study based on bosonization in 2001 \cite{orignac2001meissner}, the Meissner and vortex phases, which are reminiscent of a type-II superconductor, were found in the two-leg flux-ladder model. A variety of emergent quantum phases are hosted by the bosonic flux ladders in presence of on-site interaction \cite{calvanesestrinati2019emergent, calvanesestrinati2019pretopological, dhar2013chiral, didio2015persisting, greschner2015spontaneous, greschner2016symmetrybroken, greschner2017vortexhole, halati2023bosehubbard, johnstone2022interacting, li2021solitary, liang2021collective, orignac2001meissner, petrescu2013bosonic, petrescu2015chiral, piraud2015vortex, qiao2021quantum, song2020quantum, wei2014theory}, including Meissner phase, vortex-liquid phase, vortex lattice phase, the charge-density wave (CDW) phase, and the biased-ladder phase (BLP). The local current configurations serve as one of the most important quantities to distinguish the phases. The Meissner phase possesses a finite uniform chiral leg current encircling the ladder while the rung current vanishes. Currents in CDW and BLP are very similar with that in the Meissner phase but would break some discrete symmetries \cite{greschner2016symmetrybroken}. In the vortex phases, the rung currents on the inner rungs develop, forming vortices in the system depressing the chiral current. Except CDW, other phases can be further divided into two phases, superfluid and Mott insulator, which can be characterized by calculating the central charge, as well as the entropy \cite{buser2020interacting,greschner2016symmetrybroken}. More interestingly, the main characteristics of BLP turns out to be the finite leg-population imbalance, stabilized by the inter-chain interactions \cite{buser2020interacting}. In the thermodynamic limit the ground state would thus be twofold degenerate, and the subsequent $Z_{2}$ symmetry in terms of inversion of two legs and the signs of relevant flux is spontaneously broken \cite{greschner2016symmetrybroken}. Considering these remarkable features lead to exotic dynamical effects, therefore, we focus on the superfluid BLP phase in the present work.

\begin{figure}[h]
\includegraphics[width = 0.47\textwidth ]{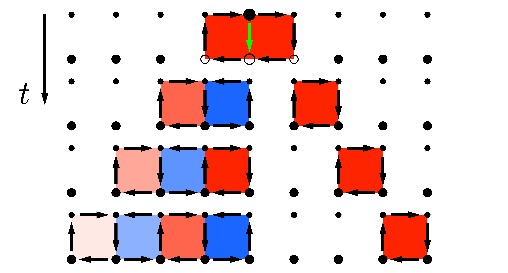}
\caption{\label{fig:1}
Sketch of the rung current excitation creating asymmetric spreading process. Typical states on the ladder at four instant points are drawn from top to bottom. At $t=0$, the rung current excitation operator (green arrow) is applied to the BLP ground state. The local population is denoted by the size of dots and the exhausted population is denoted by empty circles. The current deviated from background chiral current (not shown) is represented by black arrows. The red and blue squares label the clockwise and counterclockwise vertices, respectively, and the color depth indicates the magnitude of the vortex current.
}
\end{figure}

As sketched in Fig.\ref{fig:1}, the Hamiltonian of a paradigmatic two-leg flux-ladder model is given by
\begin{equation}
    \begin{aligned}
    H= & -J \sum_{r=1}^{L-1} \sum_{\ell=1}^2\left(a_{\ell, r}^{\dagger}    a_{\ell, r+1}+\text { H.c. }\right)\\
       &-J_{\perp} \sum_{r=1}^{L} \left( e^{-i r \phi}a_{1, r}^{\dagger} a_{2, r}+\text { H.c. }\right) \\
       & + \frac{U}{2}   \sum_{r=1}^{L}  \sum_{\ell=1}^2  n_{\ell, r}\left(n_{\ell, r}-1\right)+V \sum_{r=1}^{L} n_{1, r} n_{2, r},
    \end{aligned}
    \label{eq:1}
\end{equation}
which is residing on a ladder with $L$ rungs. Herein, the local operator $a_{\ell, r}^{\dagger} (a_{\ell, r})$ creates (annihilates) a boson on the lower $(\ell=1)$ or the upper $(\ell=2)$ leg of  the $r$-th rung; $J$ and $J_{\perp}$ are nearest-neighbor hopping constants along legs and rungs, respectively; $\phi$ is the flux per ladder plaquette; the bosons also have on-site repulsion $U$ and inter-chain repulsion $V$. Throughout this work, we set $\phi=0.85 \pi$, $J=2$, $J_{\perp}=3$, $U=2$, $\rho=0.8$, the lattice constant $a=1$ and $\hbar=1$. By this setting, the system has been determined to reside in the superfluid BLP phase \cite{greschner2016symmetrybroken}, which is essential to produce the symmetry breaking as discussed below, and we always choose the ground state with lower leg population larger than upper.

It is worth noting that, the model possesses a gauge freedom that one can choose different Peierls phase factors as long as the total flux of a single ladder plaquette remains invariant. The chosen gauge described by Hamiltonian (\ref{eq:1}) is so-called rung gauge in which the hopping matrix elements on the legs are real and on the rungs are complex.

By the Heisenberg equation of motion $ \frac{d n_{\ell, r}}{d t}  =i\left[H,n_{\ell, r}\right]$, we define a local current operator on legs as
\begin{equation}
    j_{\ell, r}^{\|}=i J\left(a_{\ell, r+1}^{\dagger} a_{\ell, r}-a_{\ell, r}^{\dagger} a_{\ell, r+1}\right),
    \label{eq:2}
\end{equation}
and the current operator on the rung as
\begin{equation}
    j_r^{\perp}=i J_{\perp}\left(e^{-i r \phi} a_{1, r}^{\dagger} a_{2, r}-e^{i r \phi} a_{2, r}^{\dagger} a_{1, r}\right).
    \label{eq:3}
\end{equation}
It is also convenient to define a background chiral current
\begin{equation}
    j^{\rm ch} = \frac{1}{L-1} \sum_{r=1}^{L-1}  \left( j_{ 1, r}^{\|} -j_{2, r}^{\|} \right)
    \label{eq:4}
\end{equation}
to characterize the average current circulating the ladder along the legs.

The ground state results are calculated by using the density matrix renormalization group (DMRG) method \cite{schollwock2011density,itensor,2008,White1993}. We simulate the flux ladder up to $L=200$ rungs and the bond dimension is typically up to 1000. As the existence of repulsive interactions, in the numerical calculations it is safe to restrict the local basis to small values. We keep at most four bosons per site and have also checked that with six bosons per site which produces consistent results. The subsequent time evolution is then simulated via time evolving block decimation (TEBD) method \cite{feiguin2011density,Vidal2007,White2004}. We typically employ bond dimension of 1000 to ensure convergence of local observables and the bond dimension up to 5000 is used in the calculation of entropy.

Let us first describe the main scenario we are investigating as displayed in Fig.\ref{fig:1}. The BLP ground state initially hosts a finite uniform anticlockwise chiral leg current encircling the ladder and vanishing rung current. A single rung in the middle is then excited to generate vortex current at time $t=0$ by applying a rung current operator without phase factor to the superfluid BLP ground state, i.e., $i \left( a_{1, L/2}^{\dagger} a_{2, L/2} - a_{2, L/2}^{\dagger} a_{1, L/2}\right) \left | \psi_{0}  \right \rangle $. The excitation creates accumulation of populations on the upper leg ($\ell = 2$) which we call it `particle' (bigger dots) and three sites with exhausting populations on the lower leg ($\ell = 1$) as we call `holes' (circles). This leads to current change on top of background chiral currents which generates two adjacent clockwise vortices (red squares) with current (black arrow) on the edge. On the right side, the `particle' moves along the same direction with the upper leg current of clockwise vortex making them constructive and move outwards like a solitary wave. On the left side, however, the concentration of `hole' and `particle' oscillate between two neighboring sites on the same rung, which is similar to the recombination process of majority and minority carriers in semiconductors. The corresponding rung current oscillates between up and down legs in the same fashion, giving rise to a vortex on the initial position alternating between clockwise and counterclockwise. In addition, vertices with opposite directions are destructive, so they move forward following with quick decay. This resembles how plane waves spread. In short, vortex currents behave particle-like on the right and wave-like on the left in just one ladder.

\begin{figure}[h]
\includegraphics[width = 0.47\textwidth ]{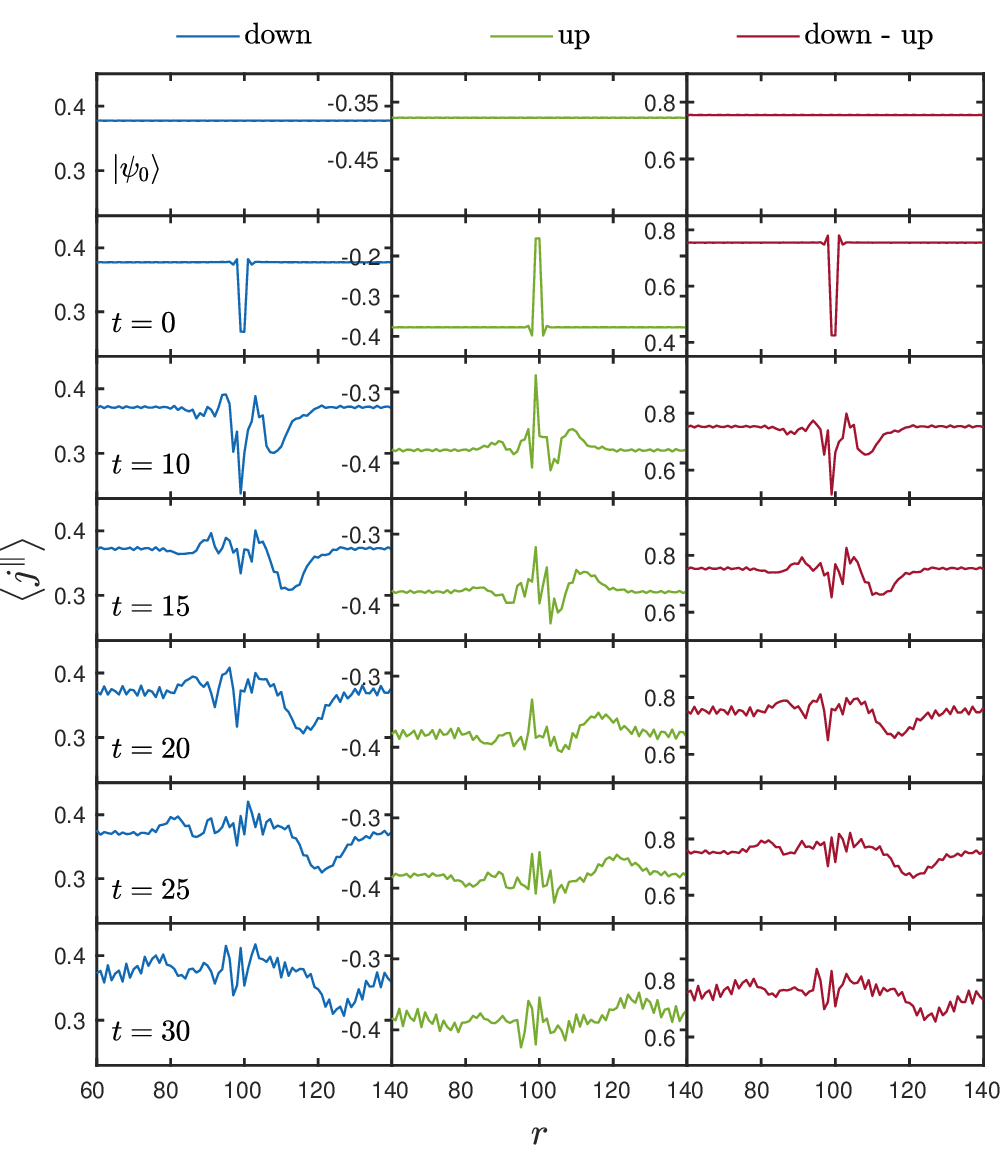}
\caption{\label{fig:2}
Snapshots of the time evolution of local leg current configurations $\left \langle j^{\|}_{\ell,r} \right \rangle$ with $V=3$ after the initial rung current excitation. `Charge' current $\left \langle j^{\rm c}_{r} \right \rangle$ is shown in the rightmost column, whose deviation from mean values quantifies the leg current of the vortices.
}
\end{figure}

Snapshots of current configurations in the time evolution after rung current excitation are shown in Fig.\ref{fig:2}. The vortex currents generated by the initial excitation on the upper leg $\left \langle j_{2,r}^{\|}\right \rangle$ and lower leg $\left \langle j_{1,r}^{\|}\right \rangle$ can be regarded as `particle' and `hole' current, respectively. The initial single wave packets of both `particle' and `hole' then broaden, which as expected by particle-hole symmetry are shown to be almost identical in magnitude but the signs are different. This means `particles' go left and `holes' go right, implying a net charge current appears. Hence, in order to manifest the symmetry breaking of spatial inversion, we define a `charge' current $\left \langle j_{r}^{\rm c} \right \rangle =\left \langle j^{\|}_{1,r}-j^{\|}_{2,r} \right \rangle$ to fingerprint the vertices. It is observed that, the central `charge' current behaves like a source, emitting vertices outward and moving. Positive and negative vortices alternate and oscillate on the left side, while a big solitary wavepacket visibly moves on the right side: The asymmetry appears.

\begin{figure}[h]
\includegraphics[width = 0.4\textwidth ]{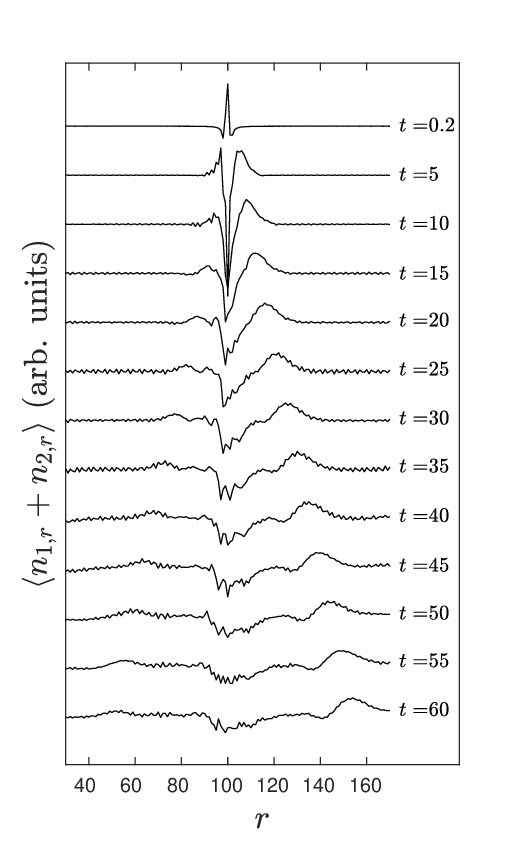}
\caption{\label{fig:3}
Snapshots of time evolution of the charge concentration configurations $ \left \langle n_{r}^{c} \right \rangle = \left \langle n_{1,r}+n_{2,r} \right \rangle$ with $V=3$.
}
\end{figure}

We relevantly calculate the population configurations on each leg $ \left \langle n_{\ell, r} \right \rangle$ (not shown) which symmetrically spread out from the middle site. To see the asymmetric influence of chiral current, we define the relevant `charge' concentration as $ \left \langle n_{r}^{\rm c} \right \rangle = \left \langle n_{1,r}+n_{2,r} \right \rangle$. As shown in Fig.\ref{fig:3}, after the wave packet splits into two counter-propagating parts, the main part of wavepacket on the left is blocked by the chiral current and the minor spreading part decays quickly. On the other hand, the decay of the right wavepacket is much slower. More importantly, regardless of the slight broadening, a distinguishable solitary wavepacket shape is always reserved. This visible asymmetry significantly exhibits that the spreading of `charge' is inversion symmetry broken in space. Although not shown, we can also parallel define the `spin' concentration whose spreading will be different from that of `charge' concentration and a similar effect of spin-charge separation could be observed.

\begin{figure}[h]
\includegraphics[width = 0.47\textwidth ]{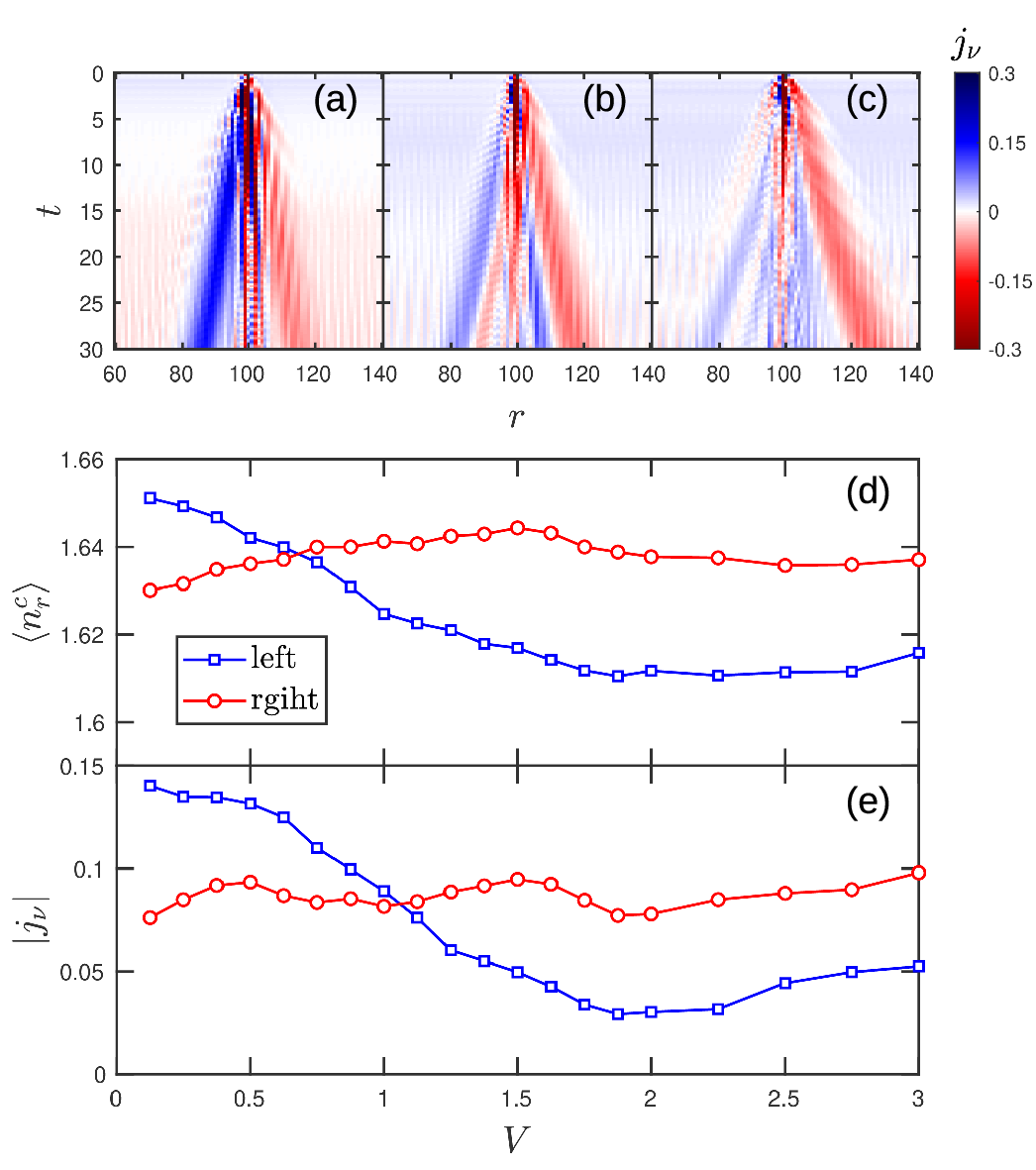}
\caption{\label{fig:4}
The net `charge' current $j_{\nu} $ and `charge' concentration $ \left \langle n_{r}^{c} \right \rangle $ in spreading of wave packets with various $V$. The two counter-propagating wave packets are shown in different colors. (a)-(c) time evolution of $j_{\nu}$ with $V=0.125$, $V=1.125$, and $V=3$. The maximum points of $ \left \langle n_{r}^{c} \right \rangle$ and $j_{\nu} $ at $t=30$ are shown in (d) and (e).
}
\end{figure}

To figure out the origin of this asymmetric spreading, the parameter of inter-chain repulsion is changed. Fig.\ref{fig:4} shows the net current $j_{\nu} = \left \langle j^{\rm c}_{r} - j^{\rm ch} \right \rangle$ of the two counter-propagating wave packets and the maximum point of `charge' concentration $\left \langle n_{r}^{c} \right \rangle$. From Fig.\ref{fig:4}(a)-(c), one can clearly see that following the increase of $V$, the wave on the left is fading while the right particle-like current is fairly stable. That is, with small $V$, the net current of the left wave is larger than the right one, and with large $V$, the left current fades and the spreading then prefers the right side. The asymmetric fading facilitates the imbalance of the two sides which mainly leads to the asymmetric spreading. In addition, the speeds on the two sides are roughly equal and linear with time.

As described above, the excitation creates the `particle' and `holes' which spread from the central rung creating a counterclockwise vortex (blue in Fig.\ref{fig:1}) moving left and a clockwise vortex (red in Fig.\ref{fig:1}) moving right. It is essential that the counterclockwise vortex enlarges the chiral current $j^{\rm ch}$ while the clockwise vertex suppresses it. For cases with small $V$, the spreading of `particle' and `hole' will weaken the imbalance of leg-population so that the system is almost close to the Meissner phase. Since the chiral current grows monotonically from BLP to Meissner phase \cite{buser2020interacting,greschner2016symmetrybroken}, the counterclockwise vortex (blue in Fig.\ref{fig:1}) will be mutually enhanced with the chiral current. Following the increase of $V$, the chiral current is suppressed and the imbalance of leg-population is then stabilized. As a consequence, the counterclockwise vortex is decreased, and when $V$ exceeds a certain value, the `particles' and `holes' on the left side are almost blocked and difficult to spread out. In this situation, the concentrations of `hole' and `particle', together with the corresponding rung current, start to oscillate on the same rung. On the other hand, considering that the clockwise vortex always decreases the chiral current, $V$ does not matter on it. Therefore, the clockwise vortex moving right overtakes, suggesting the particle-like vortex current is more robust in a counterintuitive manner.

\begin{figure}[h]
\includegraphics[width = 0.47\textwidth ]{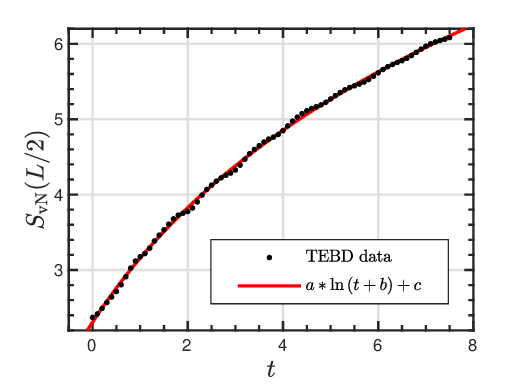}
\caption{\label{fig:5}
Time evolution of the von Neumann entanglement entropy for the central bipartition $S_{\mathrm{vN}} (L/2)$. The blue line represents the obtained curve by fitting the data via a logarithmic function $a\ln{(t+b)}+c$ with $a = 3.1829$, $b = 3.2637$, and $c = -1.4585$. The system size is $L=80$ and $V=3$.
}
\end{figure}

In a free space, the nature of microscopic particle is normally a plane wave due to the thermalization hypothesis \cite{dalessio2016quantum,deutsch2018eigenstate,srednicki1994chaos}. To comprehend the robustness of particle-like current, we further calculate time evolution of the von Neumann entanglement entropy, defined as $S_{\mathrm{vN}} = -\mathrm{Tr} [\rho \mathrm{ln} (\rho )]$ with $\rho$ being the reduced density matrix by cutting both two legs at the same points to divide the ladder into two halves, as shown in Fig.\ref{fig:5}. As the needed bond dimensions in such simulation grow exponentially with time, a smaller system size $L=80$ and a cutoff value $10^{-6}$ in the truncation are used with bond dimension up to 5000 states. One can find that as well fitted by a logarithmic function, the entropy grows logarithmically with time which is a typical signature of many-body localization \cite{abanin2017recent,abanin2019colloquium,altman2015universal,gopalakrishnan2020dynamics,Nandkishore2015}. This then interprets why the particle-like shape can be stable.

In summary, we have studied the time evolution of vortex current and concentration in a superfluid BLP phase. An asymmetric spreading of the vortex current is found determined by the direction of background chiral current. We find the right-moving particle-like current is more robust than the left-moving wave-like one, which stems from the many-body localization as indicated by the logarithm of entanglement entropy. The particle-like current therefore has got great potential in applications, such as quantum computations.


\section*{Acknowledgment}

The authors gratefully acknowledge support from the Key Research and Development Project of Guangdong Province (Grant No.~2020B0303300001), National Natural Science Foundation of China (Grant Nos.~11974118).

\bibliography{flux_ladder-v8.bbl}

\end{document}